\documentclass[aps,
 reprint,
superscriptaddress,
 amsmath,amssymb,
prb,
floatfix,
longbibliography
]{revtex4-2}
\usepackage{csquotes}
\usepackage[utf8]{inputenc}
\usepackage{graphicx}
\usepackage{dcolumn}
\usepackage{bm}
\usepackage{xcolor}
\usepackage{microtype}
\usepackage[
    separate-uncertainty = true,
    ]{siunitx}
\usepackage{cleveref}

\begin{document}

\title{Laser-induced metastable (anti-)skyrmion states with higher-order topological charge}

\author{Tim Titze}
\affiliation{I.\,Physikalisches Institut, Universit\"at G\"ottingen, 37077 G\"ottingen, Germany\looseness=-1}
\author{Marcel Möller}
\affiliation{Department of Ultrafast Dynamics, Max Planck Institute for Multidisciplinary Sciences, G\"ottingen, Germany\looseness=-1}
\author{Timo Schmidt}
\affiliation{Institute of Physics, University of Augsburg, 86135 Augsburg, Germany\looseness=-1}
\author{Mariam Hassan}
\affiliation{Institute of Physics, University of Augsburg, 86135 Augsburg, Germany\looseness=-1}
\author{Sabri Koraltan}
\affiliation{Institute of Applied Physics, TU Wien, Vienna, A-1040, Austria}
\author{Stefan Mathias}
\affiliation{I.\,Physikalisches Institut, Universit\"at G\"ottingen, 37077 G\"ottingen, Germany\looseness=-1}
\affiliation{International Center for Advanced Studies of Energy Conversion (ICASEC), Universit\"at G\"ottingen, 
37077 G\"ottingen, Germany} 
\author{Manfred Albrecht}
\affiliation{Institute of Physics, University of Augsburg, 86135 Augsburg, Germany\looseness=-1}
\author{Claus Ropers}
\email[]{claus.ropers@mpinat.mpg.de}
\affiliation{Department of Ultrafast Dynamics, Max Planck Institute for Multidisciplinary Sciences, G\"ottingen, Germany\looseness=-1}
\affiliation{IV.\,Physikalisches Institut, Universit\"at G\"ottingen, 37077 G\"ottingen, Germany}
\affiliation{International Center for Advanced Studies of Energy Conversion (ICASEC), Universit\"at G\"ottingen, 37077 G\"ottingen, Germany}
\author{Daniel Steil}
\email[]{dsteil@gwdg.de}
\affiliation{I.\,Physikalisches Institut, Universit\"at G\"ottingen, 37077 G\"ottingen, Germany\looseness=-1}

\begin{abstract}
    Ultrafast creation and manipulation of topologically non-trivial spin objects promises significant potential for spintronic applications. Here, we demonstrate laser-induced nucleation of skyrmions and antiskyrmions with topological charge up to $|Q|=3$ from the saturated magnetization state of a Co/Ni-multilayer in which these textures are absent without optical excitation. The resulting configuration of spin objects thus constitutes a metastable, thermodynamically hidden magnetic state. We control the number and topological-charge distribution of spin objects by the excitation fluence and the external out-of-plane magnetic field.   
\end{abstract}

\maketitle

\section{Introduction}
Since their discovery in 2009 by Mühlbauer et al.~\cite{Muehlbauer2009}, magnetic skyrmions~\cite{Bogdanov1994, Roessler2006} have attracted significant research interest due to their tremendous potential for future spintronic and magnonic applications. Potential applications range from racetrack memory~\cite{Fert2013,Tomasello2014}, logic devices~\cite{Zhang2015,Finocchio2016,Fert2017}, and microwave components~\cite{Wintz2016,Yu2021,Petti2022,Koraltan2024} to the emerging field of unconventional computing~\cite{Huang2017,Song2020,Li2021,Yokouchi2022}. All of these concepts require the ability to detect and manipulate (e.g., shifting, nucleating and annihilating) magnetic skyrmions, as well as harnessing their  dynamic properties. 

Metallic multilayers are readily fabricated and scalable platforms for such applications, where skyrmionic spin objects are stabilized either by the interfacial Dzyaloshinskii–Moriya interaction (DMI)~\cite{Chen2015,Woo2016,MoreauLuchaire2016,Jaiswal2017,Soumyanarayanan2017} or by the interplay between dipolar interactions and perpendicular magnetic anisotropy even above room temperature~\cite{Montoya2017a,Montoya2017b,Heigl2021,Hassan2024,Titze2024,Titze2025}. Recently, Hassan and coworkers reported the existence of high-order (anti)skyrmionic spin objects in Co/Ni multilayer films, which are field-nucleated from domains walls containing vertical Bloch lines by applying an out-of-plane (oop) magnetic field~\cite{Hassan2024}. These spin objects are predicted to exhibit a significantly reduced skyrmion Hall effect, opening up new pathways in the direction of novel applications. 

\begin{figure}[ht]    
    \centering
    \includegraphics[width=1\columnwidth]{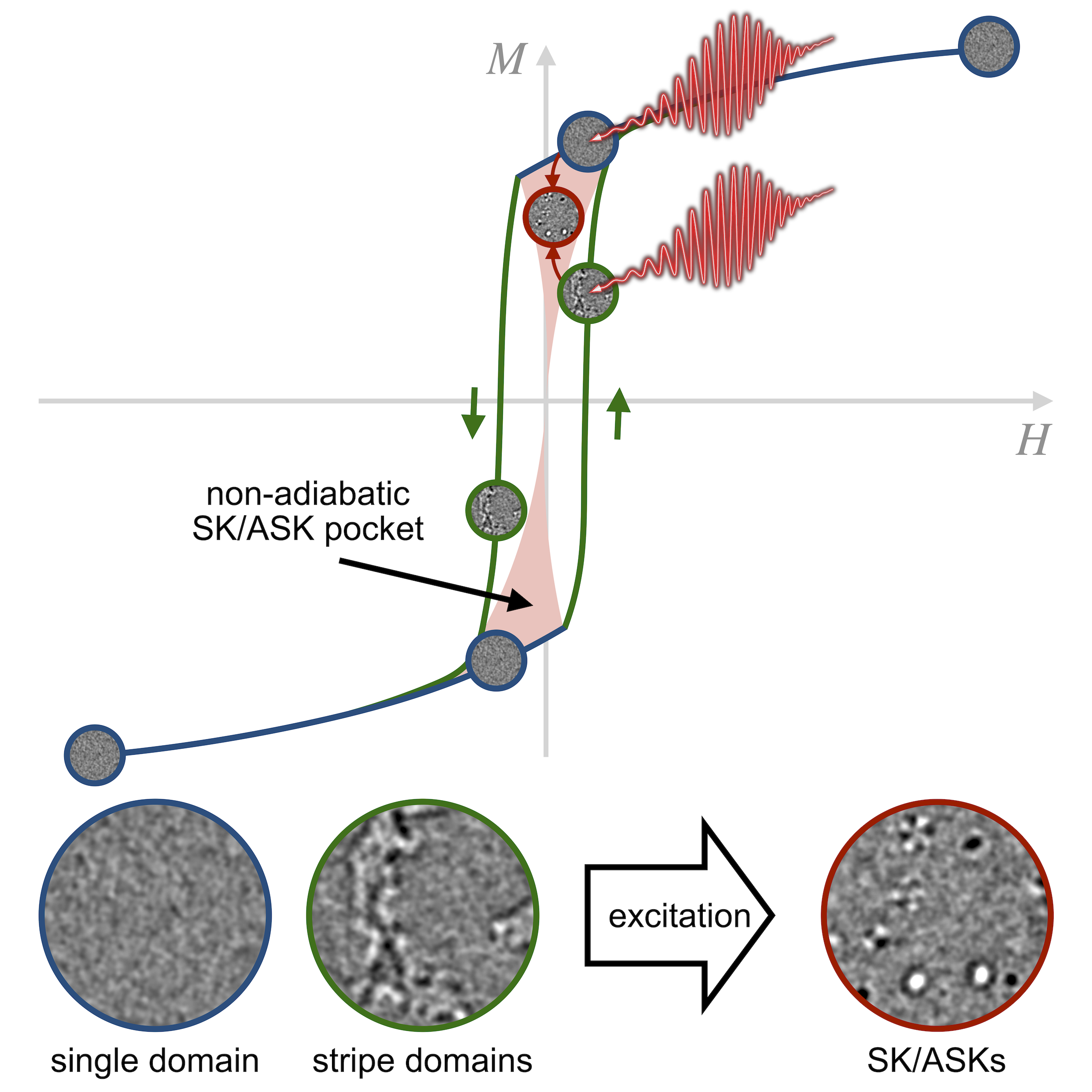}
     \caption{Illustration of laser-induced nucleation of disordered, hidden skyrmions/antiskyrmions (red circle) starting from either a single domain (blue) or a stripe domain state (green).}
     \label{fig:Intro}
\end{figure}
\begin{figure*}[hbt]
     \centering
     \includegraphics[width=1\textwidth]{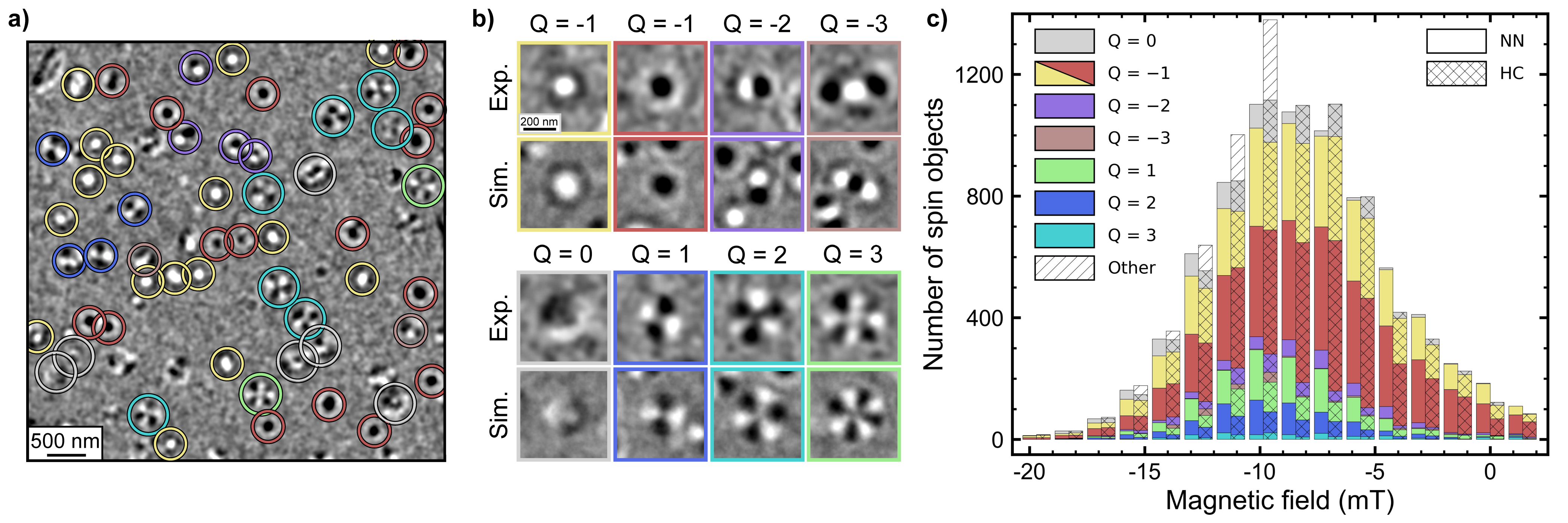}
     \caption{a) Various laser-nucleated spin objects encircled with colors in an example LTEM image with parameters $\mu_0H=-11.3$\,mT and $F=4.8$\,mJ/cm$^2$. b) Selected laser-induced spin objects with different topological charge $Q$ imaged by LTEM compared to calculated LTEM contrast (not to scale) from micromagnetic simulations at a field of $\mu_0H=-28$\,mT (see Appendix~\ref{App:C}). c) Bars show the magnetic field-dependent number of nucleated spin objects by applying a laser fluence $F=4.8$\,mJ/cm$^2$ coming from an initially negatively saturated state. Plain bars correspond to data evaluated using a neural network (NN). Crosshatched bars correspond to manually evaluated (hand counted, HC) numbers of identified spin objects as a comparison. Bar colors correspond to topological charges $Q$ as depicted in b). Note that the colors yellow and red correspond to clockwise and counterclockwise skyrmions, respectively. The diagonal-hatched\enquote{Other}-category comprises spin objects which could not be conclusively assigned. Insufficent object separation prevents counting such objects for fields larger than $\mu_0H=-9.9$\,mT.}
     \label{fig:Nucleation}
\end{figure*}
For the efficient manipulation of such skyrmionic objects using spin transfer or spin orbit torques~\cite{Ralph2008, Manchon2019}, very thin Co/Ni magnetic multilayers would be needed, where, however, no such objects appear in case of magnetic field-cycling at ambient temperature. In this work, we demonstrate laser-induced nucleation of spin objects with higher-order topological charge in a Co/Ni multilayer which does not intrinsically host a skyrmionic field-induced state. Following single-pulse laser excitation in a Lorentz transmission electron microscopy (LTEM) with \textit{in-situ} optical excitation, we observe a distribution of nucleated high-order skyrmions and antiskyrmions (SK/ASKs) created either from a saturated magnetization state, or from a stripe domain state, as schematically depicted in Fig.~\ref{fig:Intro}. We systematically analyze the conditions facilitating the nucleation of these spin objects by varying both the external magnetic field and the laser fluence. In particular, we find that the excitation strength controls the relative occurrence of objects with different topological charge. The demonstration of laser-nucleated metastable, thermodynamically hidden spin objects with topological charge $|Q|>1$ constitutes a significant step towards deterministic light-induced generation of complex spin textures for next-generation spintronics devices.

\section{Results and discussion}
A sputtered [Co($0.2$\,nm)/Ni($0.7$\,nm)]$_6$ multilayer sample exhibiting perpendicular magnetic anisotropy is imaged using an LTEM with \textit{in-situ} optical excitation. Single laser pulses with a pulse duration of approximately $300$\,fs at a center wavelength of $750$\,nm are used to excite the sample at near normal incidence. The measurements are performed for variable laser fluence and for different out-of-plane (oop) magnetic fields (see also Fig.~\ref{fig:Intro}). Further information on the setup, film preparation, and magnetic film properties are provided in Appendix~\ref{App:C}. 

\subsection{Type and magnetic field-dependent distribution of spin objects}
Figure~\ref{fig:Nucleation}a highlights the diversity of topological spin objects created upon laser-excitation in LTEM for a magnetic field of $\mu_0H=-11.3$\,mT and a laser fluence of $F=4.8$\,mJ/cm$^2$. We categorize the SK/ASKs based on their topological charge $Q$, with the typical LTEM contrasts depicted in Fig.~\ref{fig:Nucleation}b for $Q=0,\,\pm 1,\,\pm 2,\, \pm3$ together with LTEM contrast of such objects calculated using data from micromagnetic simulations. The colored circles around the objects correspond to their topological charge given in Fig.~\ref{fig:Nucleation}b. In addition, we also find rare instances of SK/ASKs with even higher $Q$, which are not categorized (see Appendix~\ref{App:B}, Fig.~\ref{fig:HOASK}). 

\begin{figure*}[ht]
     \centering
     \includegraphics[width=\textwidth]{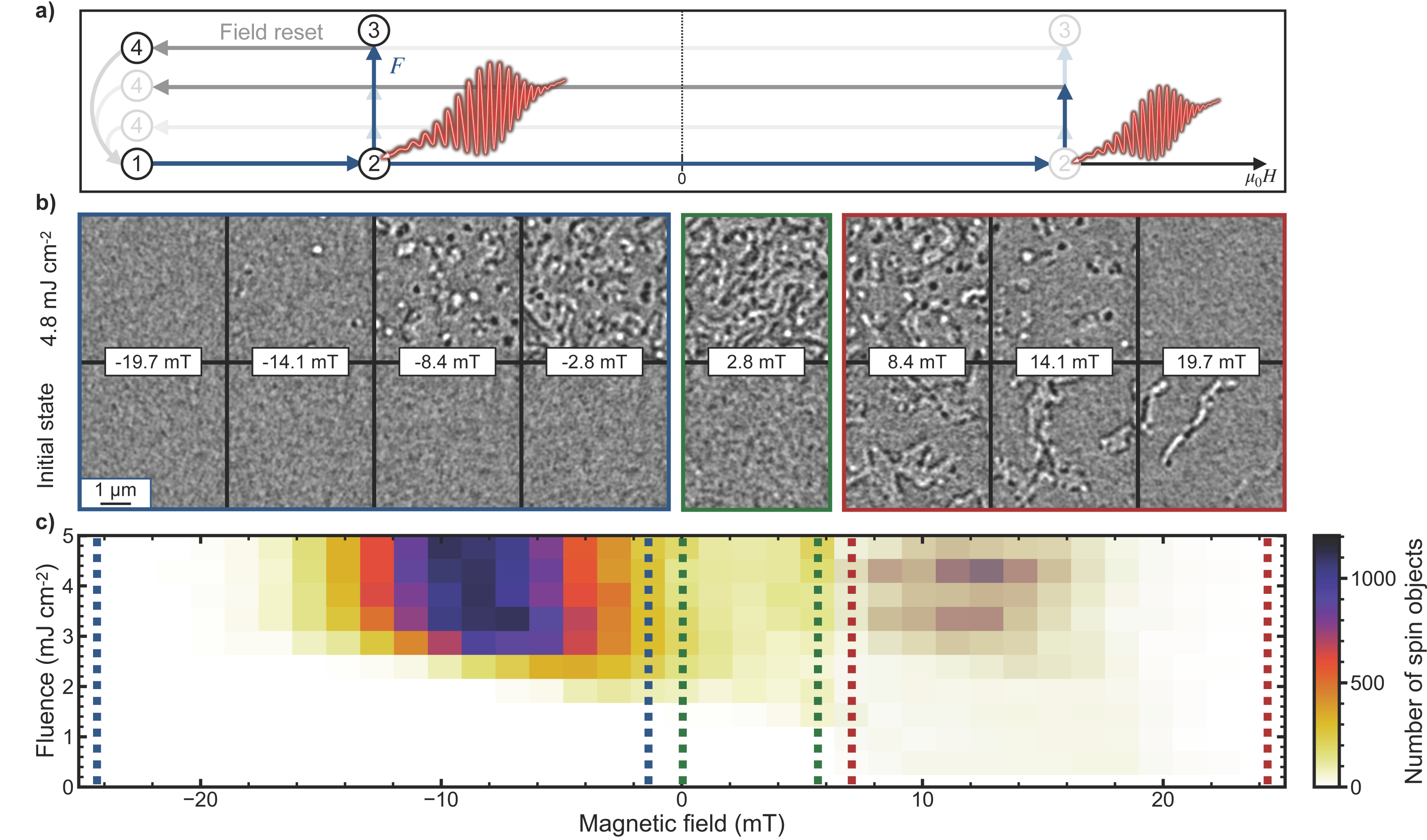}
     \caption{a) Protocol for measuring the magnetic field-dependent laser-induced generation of topological spin textures for different fluences. From negative saturation (1) a certain oop magnetic field is set (2) and a first LTEM image is taken, the sample is laser-excited (3), and a second LTEM image is taken. Afterwards (4), the sample is reset by going again to negative saturation and the procedure is repeated for different fluences $F$. b) Magnetic field-dependent spin textures for the field-prepared initial state (lower panel) and after single-pulse laser excitation with fluence $F=4.8$\,mJ/cm$^2$ (upper panel) corresponding to the protocol depicted in a). c) Total number of spin objects for different combinations of fluence and magnetic field counted by the neural network. The regions between the dashed lines mark the different magnetic field-prepared states before laser excitation: single domain (blue), transition state (green), and stripe domains (red). Objects numbers in the stripe domain regions (grayscale shading) are qualitative only, see text.}
     \label{fig:Overview}
\end{figure*}

To efficiently evaluate the large number of laser-nucleated spin objects in an LTEM image of size $21.4\times21.4$\,{\textmu}m$^2$, we trained a neural network for classification and counting (see Appendix~\ref{App:C}). We then used this neural network to study the dependence of laser-induced SK/ASK nucleation on magnetic field and excitation fluence, limiting the search to the $Q$-values given in Fig.~\ref{fig:Nucleation}b. This analysis yields the magnetic field-dependent number of different spin objects nucleated from an initially negatively saturated state at a fluence of $F=4.8$\,mJ/cm$^{2}$, displayed in Fig.~\ref{fig:Nucleation}c (open bars). In addition, for the data depicted in Fig.~\ref{fig:Nucleation}c, manual counting was employed to analyze the performance of the neural network (crosshatched bars). In both cases we find that skyrmions with topological charge $Q=-1$ are most common, while higher-order SK/ASKs only amount to at maximum about $1/4$ of the total objects. The observation of larger percentages of higher-order SK/ASKs is mostly restricted to intermediate magnetic fields in the vicinity of  $\mu_0H=-11.3$\,mT, whereas $Q=-1$ skyrmions and $Q=0$ bubbles can be stabilized over a broader field range. We believe that this difference stems from the distinct origin of the higher-order SK/ASKs, which require vertical Bloch lines for their nucleation, as discussed for ground state spin objects in thicker Co/Ni multilayers~\cite{Hassan2024}. Such topological defects do not exist in the negatively saturated magnetic state, indicating that they transiently form after laser excitation in a magnetic energy landscape controlled by the out‑of‑plane magnetic field. 

Comparing the performance of the neural network with the hand-counted data in more detail, it is evident that the neural network generally detects more higher-order spin objects, except for the case of triskyrmions ($Q=-3$), whereas human analysis typically detects a higher number of trivial type-II bubbles ($Q=0$). Furthermore, the neural network does not count objects it cannot identify, whereas this is possible for hand-counting and performed here up to $\mu_0H=-9.9$\,mT. For more positive magnetic field values, the number of such unidentified spin objects increases even further---at least initially---and counting them becomes too challenging to report reliable numbers~\footnote{For higher object densities, overlapping of such spin objects in the defocused LTEM images makes number determination increasingly inaccurate.}. Part of the objects classified as \enquote{Other} are higher order spin objects, which could not be accurately classified. Overall, the spin-object counts obtained by neural-network analysis and human evaluation exhibit a similar dependence on magnetic field. Absolute values in both cases, however, depend on the detection method.

\subsection{Stability regions for metastable SK/ASK nucleation}

We next investigate how SK/ASK nucleation depends on the two external control parameters magnetic field and laser-excitation fluence using the previously trained neural network. All measurements are performed starting from negative magnetic saturation at specific magnetic field values and fluences, as explained by the protocol displayed in Fig.~\ref{fig:Overview}a. Note that the sample is not homogenously illuminated by the laser beam, i.e., we observe a gradient in excitation across the images. 

In the field-dependent magnetic state without laser excitation, hardly any skyrmionic spin objects are observed, as depicted in the lower panel of Fig.~\ref{fig:Overview}b. In particular, no skyrmionic spin objects are evident in the negative magnetic field range coming from negative saturation. This observation is explained by the comparatively low thickness of our sample ($5.4$\,nm), which is below the threshold required to stabilize topological spin textures in the field-prepared initial state through the balance of dipolar interactions and perpendicular magnetic anisotropy~\cite{Hassan2024}. In more detail, starting from a saturated single domain state at large negative magnetic field, the system persists in this state upon decreasing the magnetic field up to zero. Further field increase leads to the emergence of stripe domains appearing to be stable within a magnetic field range of $\mu_0H=7$--$20$\,mT. No significant amount of distinct and well-separated spin objects can be observed in this parameter range, only few skyrmionic objects can be found.

Ultrafast optical excitation of these field-prepared states with fluence $F=4.8$\,mJ/cm$^2$ leads to the nucleation of a significant number of metastable high order SK/ASKs, as visible in large parts of the upper panel of Fig.~\ref{fig:Overview}b. These are created from both the single-domain and the stripe-domain state, as long as the combination of laser fluence and external magnetic field matches, similar to the laser-induced nucleation of $Q=-1$ skyrmions reported in studies on Co/Pt-based multilayers~\cite{Buettner2021,Gerlinger2021}. However, in contrast to Co/Pt multilayers, in the Co/Ni system strong interfacial DMI associated with heavy-element spin–orbit coupling is absent. Ferrimagnetic Fe/Gd-based multilayers represent a more closely related system, hosting dipolar-stabilized (anti)skyrmionic spin objects~\cite{Lee2016FeGd,Montoya2017a,Montoya2017b,Desautels2019,Heigl2021,Titze2024,Titze2024b,Titze2025, Titze2026}. Laser-induced nucleation of skyrmions has also been observed in this system, yet only from the stripe domain state~\cite{Titze2024}, likely due to inhomogeneous sample excitation as the sample thickness is larger than the laser penetration depth in this case~\cite{Titze2025}. 

Figure~\ref{fig:Overview}c now showcases how the number and stability of nucleated spin objects is controlled by both magnetic field and excitation fluence. In general, a higher fluence increases the number of nucleated spin objects and slightly broadens the magnetic field range that allows for SK/ASK nucleation. However, there exist fluence and magnetic field combinations for which few or even no SK/ASKs are created. Nucleation from the single domain state (region marked by blue dashed lines) requires a threshold fluence $F_{\mathrm{min}}=1.9$\,mJ/cm$^2$, which increases with increasing magnitude of the magnetic field. In the transition region from the single domain state to the stripe domain state, the field range $\mu_0 H=0$--$7$\,mT (region marked by green dashed lines), primarily stripe domains and only a small number of SK/ASKs are generated, see the snapshots at $\mu_0 H=2.8$\,mT in Fig.~\ref{fig:Overview}b. Excitation of a stripe domain ground state at $\mu_0H>7$\,mT (region marked by red dashed lines and shaded in grayscale) again leads to the formation of an SK/ASK state with a large amount of topological spin objects in the field range $\mu_0H\approx 8$--$16$\,mT for fluences $F>3.0$\,mJ/cm$^2$. 

The presence of stripe domains in the field-prepared state does, however, inhibit reliable object classification and counting using the neural network, as a background subtraction using the LTEM image before laser illumination for the same magnetic field is not possible (for details see Appendix~\ref{App:C}). This lowers the detection efficiency of the neural network, i.e., spin object numbers are typically underestimated and object designation is less reliable. For this reason, data in this region should be understood as depicting trends only~\footnote{Visual inspection of LTEM images confirms large numbers of spin objects for magnetic fields in the range $\mu_0H\approx 8$--$16$\,mT for fluences $F\geq 2.5$\,mJ/cm$^2$.}. A qualitative comparison still reveals, that the maximum number of spin objects seems to occur at somewhat higher magnetic field than for the negative field range. Furthermore, there is no well-defined nucleation threshold anymore, as the presence of stripe domains in the field-prepared state apparently lowers the barrier for skyrmion nucleation. Most efficient skyrmion nucleation is, however, still observed for similar fluence ranges as in the negative field case, concomitant with a near complete erasure of the spin textures present before laser illumination (see also Appendix A).

\subsection{Mechanism for nucleation of metastable SK/ASKs}

Light-induced creation of metastable topological spin objects using ultrashort light pulses has been discussed in the literature over the last 10 years for magnetic vortices, (anti)skyrmions, and compound objects like skyrmion bags and target skyrmions~\cite{Eggebrecht2017, Je2018, Berruto2018,Buettner2020, Buettner2021,Zhang2023,Khela2023,Truc2023, Kalin2024, Titze2024, Li2024, Zhu2024, Kern2025}. 

In general, one can distinguish two limiting cases for the creation of such objects in metallic ferromagnets: (i) nucleation of SK/ASKs by quasi-static laser-heating on an equilibrium path from a spin-textured state (e.g., stripe domains) into a skyrmion (lattice) state~\cite{Berruto2018, Kalin2024}, or (ii) creation by strong laser-excitation into a transiently fluctuating nonequilibrium magnetization state, where SK/ASKs may nucleate from topological defects even from an initially saturated magnetization state~\cite{Buettner2021,Li2024,Zhu2024}. In both cases, the subsequent rapid thermal quench of the magnetic layer~\cite{Eggebrecht2017, Liefferink2026} leads to the freezing-in of the metastable topological spin objects, which would otherwise not exist for the initial magnetic field and temperature settings~\cite{Titze2025}. Particularly in ferromagnetic insulators, further aspects have been discussed besides this mechanism such as the inverse Faraday effect, Raman excitation of phonon and magnon modes, and direct magnetoelectric coupling to the spontaneous electric polarization~\cite{Truc2023}.

For our metallic Co/Ni multilayers, the quasi-static pathway is unlikely, as SK/ASK nucleation in this case requires the presence of stripe domains and vertical Bloch lines~\cite{Hassan2024}, similar to the case of other dipolar-stabilized systems~\cite{Montoya2017a,Heigl2021,Titze2024, Titze2025}. Such spin structures are absent in the saturated initial state and we do not expect them to appear upon heating as the dipolar interactions ($\propto M^2_{\mathrm{sat}}$) are reduced. Thereby, we do not anticipate the existence of an thermally accessible SK/ASK state at negative magnetic fields, i.e., optical excitation of the single domain state should not induce such a state in a heat-driven nucleation process.

In contrast, nucleation via a transient topological fluctuation state is expected to be largely insensitive to the initial spin textures. We thereby assume a nucleation mechanism involving the laser-induced generation of a transiently fluctuating magnetization state~\cite{Eggebrecht2017, Buettner2021, Li2024, Liefferink2026}. 
Note that in our case, even after strong excitation, a few stripe domains persist when nucleating from the stripe domain phase. In addition, SK/ASKs predominantly nucleate in regions that were previously occupied by these stripe domains, indicating that a complete magnetization quench of the film is unlikely. This also explains the observed asymmetry between negative and positive magnetic field values, as the laser-induced formation of magnetic textures is still influenced by the initial magnetization state.

\subsection{Analysis of the energy barrier for nucleation of different spin objects}

\begin{figure}[t]
     \centering
     \includegraphics[width=1\columnwidth]{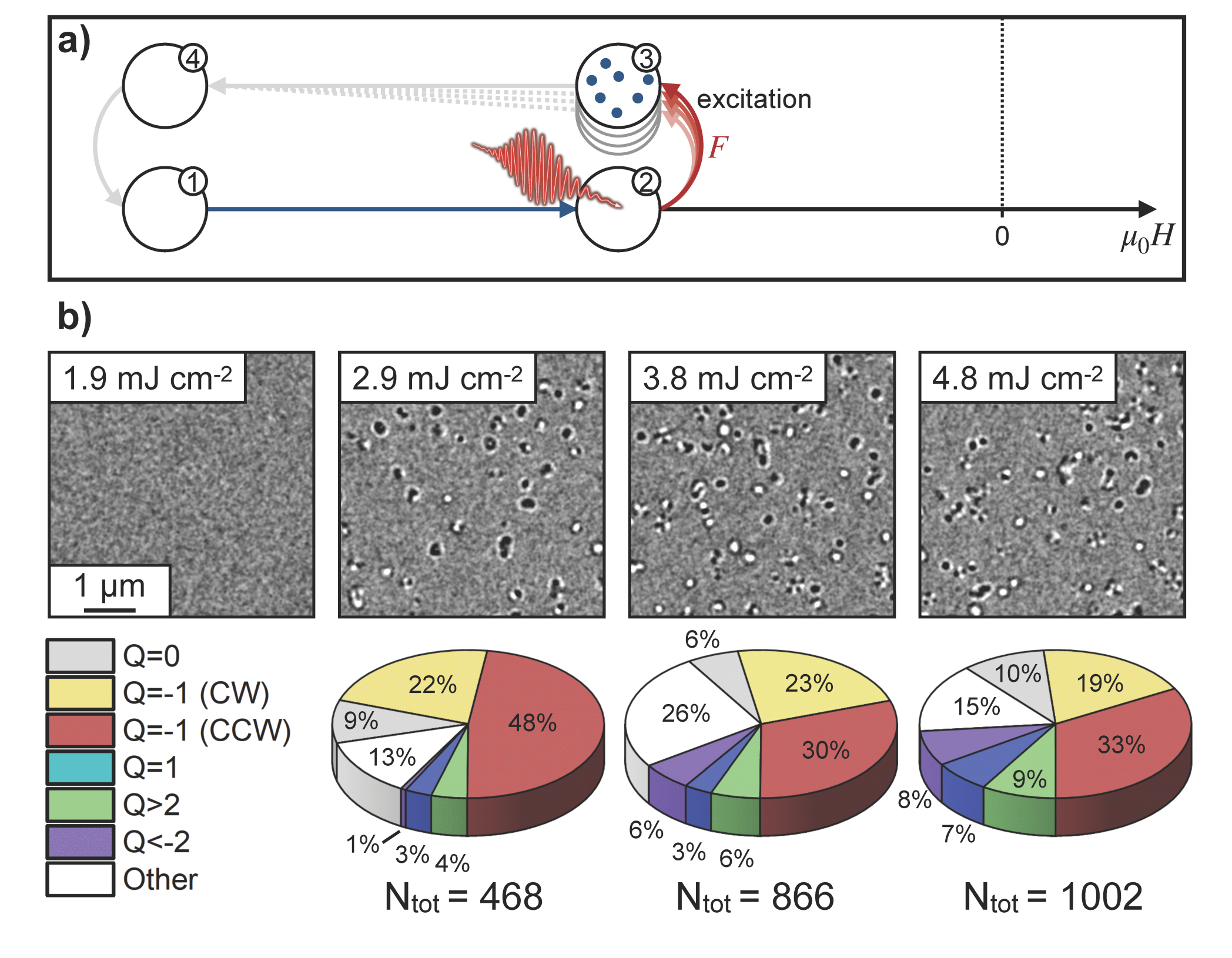}
     \caption{a) Protocol for measuring fluence-dependent nucleation of magnetic spin objects. Starting from an initial saturated state (1), the external oop magnetic field is increased to $\mu_0 H=-11.3$\,mT (2). Then, the system is excited using a single laser pulse with variable fluence (3) and an LTEM snapshot is taken. Finally, the initial saturated state is restored by resetting the external magnetic field (4). b) LTEM snapshots of nucleated spin textures using various excitation fluences at a constant magnetic field $\mu_0 H=-11.3$\,mT with corresponding pie charts illustrating the fraction of the different spin objects.}
     \label{fig:Fluence}
\end{figure}

The distribution of topological charges of laser-nucleated SK/ASKs depends not only on the magnetic field, but also on the laser excitation fluence, as already observed in Fig.~\ref{fig:Overview}. In Fig.~\ref{fig:Fluence}, we analyze how the share of nucleated spin objects at a constant negative magnetic field $\mu_0 H=-11.3$\,mT changes with the excitation fluence used for nucleation from the single domain state. As mentioned earlier, at negative magnetic fields spin objects do not nucleate below the threshold fluence $F_{\mathrm{min}}=1.9$\,mJ/cm$^2$. Laser fluences in the vicinity of this threshold lead to the generation of isolated spin objects which predominantly exhibit low topological charge. Counterclockwise (CCW) skyrmions nucleate preferably compared to clockwise (CW) skyrmions, even though both exhibit a topological charge of $Q=-1$. The origin of this chiral symmetry breaking is intrinsically not clear, as in multilayers without DMI both rotation senses should nucleate equally. However, for the LTEM sample, strain effects or even curvature caused by membrane buckling can possibly induce DMI-like effects~\cite{Zhang2021, Raftrey2025, Koraltan2025,Schmidt2026} which can give rise to this asymmetry. Increasing the fluence favors the nucleation of higher-order antiskyrmions, as both their absolute number and relative fraction rise significantly with fluence, while the number of $Q=-1$ skyrmions increases slowly and the relative fraction of these objects shrinks. In terms of the energy landscape, these results clearly demonstrate that skyrmions exhibit the lowest energy barrier for nucleation, while that of higher-order spin objects is increased~\cite{Hassan2024}. 

\section{Conclusion}
In conclusion, we demonstrate optical preparation of metastable skyrmions and antiskyrmions with higher-order topological charge using single femtosecond laser pulses. Single-shot laser excitation induces a large number of metastable skyrmionic textures---a hidden state that does not emerge in the absence of optical stimulation. We assume that a nucleation mechanism involving a transiently fluctuating magnetization state generated by the laser excitation is responsible for this. We find a multitude of skyrmionic spin objects, with topological charges regularly reaching up to $Q=\pm 3$, and in rare cases even beyond. The number of nucleated spin objects strongly depends on the combination of external oop magnetic field and laser fluence. Furthermore, our results reveal that the excitation strength affects the distribution of topological charges among the objects. These observations highlight the potential of ultrafast optical excitation as a tool for controlling hidden topological spin structures with topological charge other than unity. The nucleation and modification of such hidden states in technologically relevant metallic multilayer films at room temperature opens up promising pathways towards novel skyrmionic applications and for fundamental studies of the dynamical properties of higher-order topological spin textures. Further progress in this field may be possible using spatially or temporally structured light for topological texture nucleation to directly imprint desired textures in the creation process of topological fluctuation states~\cite{Jiang2024,Mitra2025}.

\begin{acknowledgments}
D.S. and M.A. gratefully acknowledge funding by the Deutsche Forschungsgemeinschaft (DFG, German Research Foundation) via project no.\,575043137. T.T. thanks M. Staabs for helping in training the neural network. M.M. and C.R. acknowledge funding by the DFG via resources from the Gottfried Wilhelm Leibniz Prize (RO 3936/4-1), and acknowledge support by the Göttingen Ultrafast transmission electron microscopy (UTEM) team and helpful discussions in particular with Tyler Harvey and John Gaida.
\end{acknowledgments}

M.M., C.R., and M.A. conceptualized the study. M.M. performed the LTEM measurements under supervision of C.R. Sample preparation and pre-characterization was performed by T.S. and M.H. under supervision of M.A. T.T. trained the neural network and evaluated the data under supervision of D.S. D.S. hand-counted the spin objects. S.K. performed micromagnetic simulations for different spin objects. T.T. and D.S. wrote the manuscript with input from all authors.

\subsection*{Data availability}
The data that support the findings of this study are available from the authors upon reasonable request.

\appendix
\section{SK/ASK nucleation from the stripe domain phase}\label{App:A}
\begin{figure}[h!]
     \centering
     \includegraphics[width=1\columnwidth]{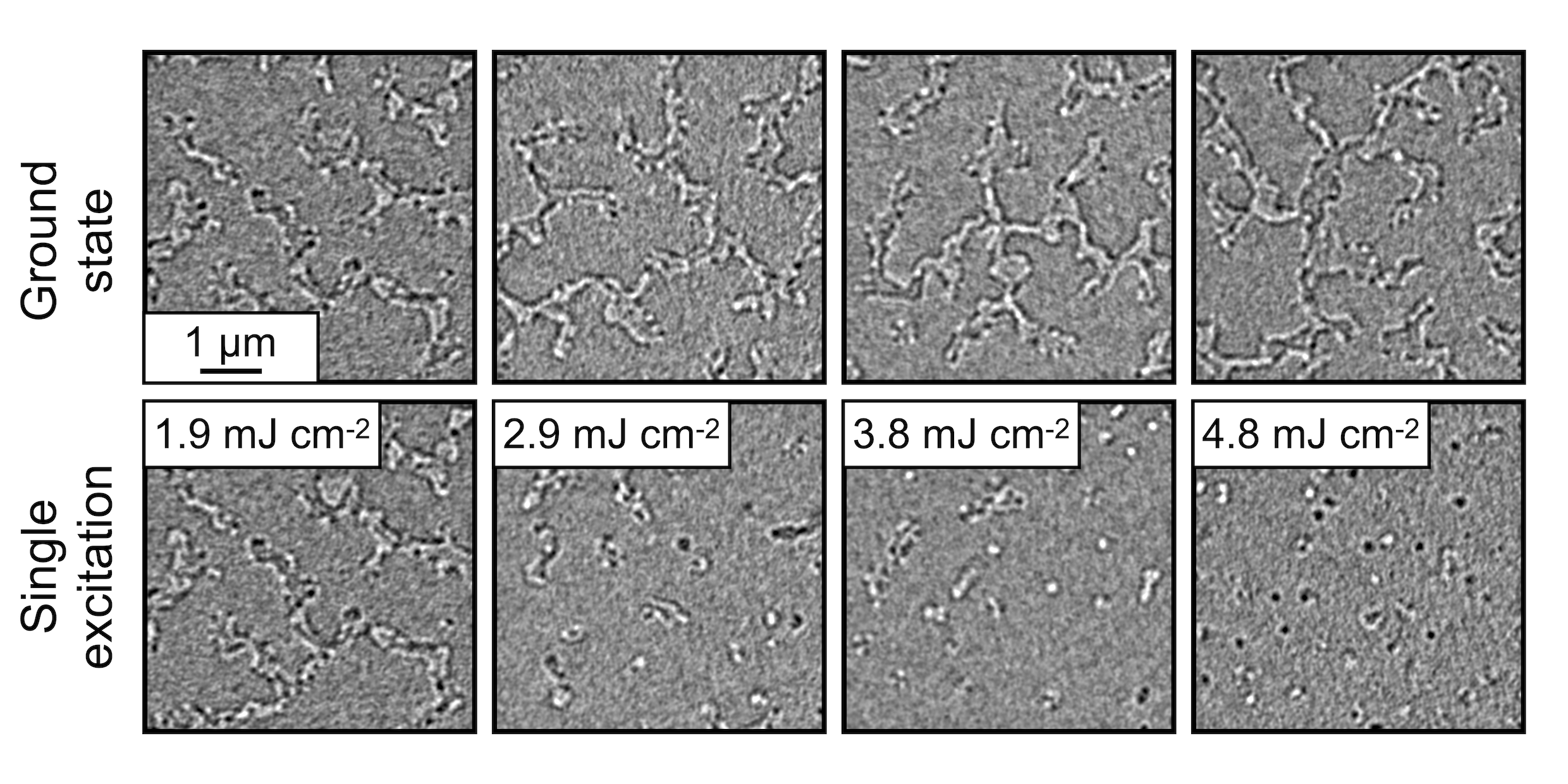}
     \caption{Fluence-dependent single-shot nucleation of SK/ASKs starting from a stripe domain phase at $\mu_0H=14.1$\,mT.}
     \label{fig:StripetoASK}
\end{figure}
Starting from a stripe domain state, i.e., positive magnetic fields, SK/ASKs predominantly nucleate in regions that were inherently occupied by stripe domains, see Fig.~\ref{fig:StripetoASK}. However, for the strongest excitation ($F=4.8$\,mJ/cm$^2$) SK/ASKs also appear in formerly saturated regions. In addition, a few SK/ASKs are already observed in the field-prepared state. A detailed neural network-based analysis of the SK/ASK quantities in the stripe domain state is difficult, as the obtained images are less suitable for analy\-sis compared to the ones at negative magnetic fields. The reason for this difference is given by the background subtraction procedure used to remove unwanted noise and image artifacts. For details refer to the Appendix~\ref{App:C} on data analysis. 

\section{High-order SK/ASKs}\label{App:B}
\begin{figure}[h!]
     \centering
     \includegraphics[width=\columnwidth]{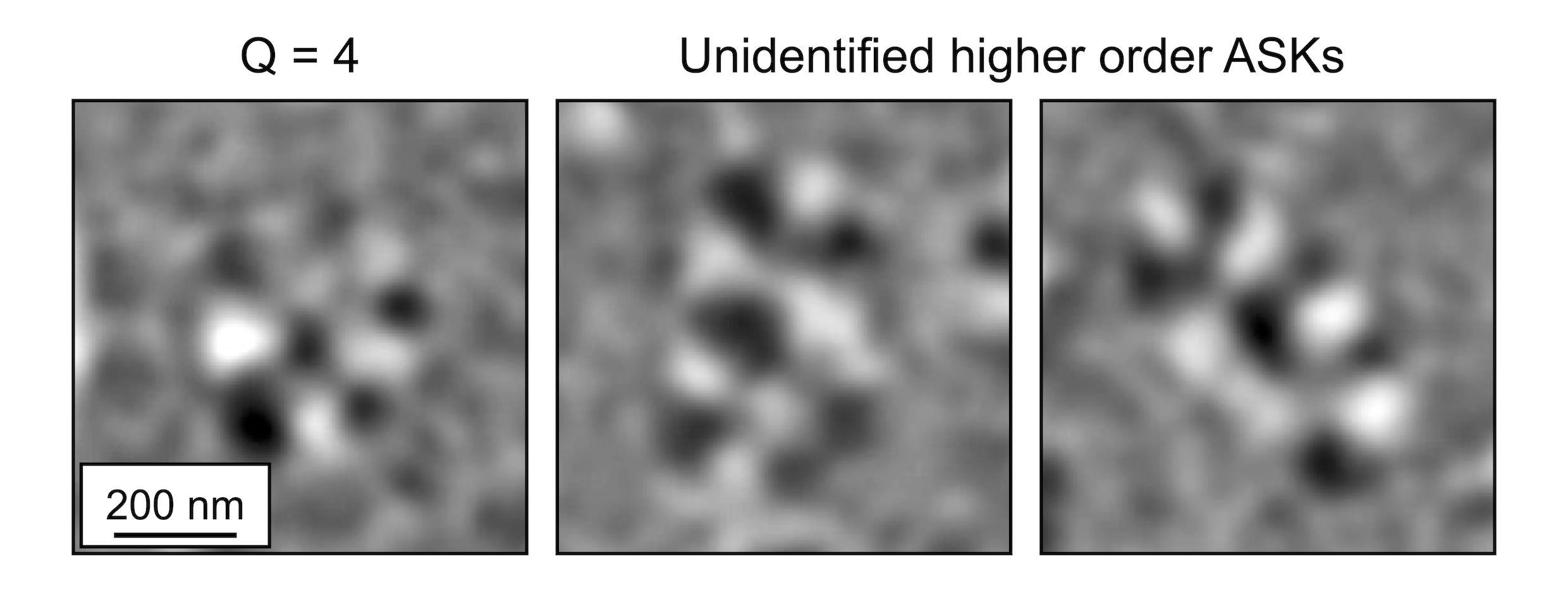}
     \caption{Rare spin objects such fourth order antiskyrmions ($Q=4$), as well as unidentified higher-order spin objects.}
     \label{fig:HOASK}
\end{figure}

In addition to the spin objects presented in Fig.~\ref{fig:Nucleation}, we also observe rarer objects which are shown in Fig.~\ref{fig:HOASK}. These include higher order antiskyrmions ($Q=4$), however, their number is so small that the neural network cannot be reliably trained onto these. Furthermore, we observe higher-order spin objects that remain unidentified, two of which are displayed in Fig.~\ref{fig:HOASK}. These could be explained by the merger of two or more SK/ASKs, similar to those presented in Ref.~\cite{Hassan2024}. 

\section{Methods}\label{App:C}

\subsection{Sample preparation and characterization}
[Co($0.2$\,nm)/Ni($0.7$\,nm)]$_6$ multilayers were prepared on both thermally oxidized Si(100) substrates and on $30$-nm-thick Si$_3$N$_4$ membranes required for LTEM imaging. The multilayer samples were deposited at room temperature by dc magnetron sputtering. The sputter process was carried out using an Ar working pressure of $3.5$\,{\textmu}bar in an ultra-high vacuum chamber. A $3$-nm-thick Pt seed layer and a $3$-nm-thick Si$_3$N$_4$ capping layer were used to protect the films from oxidation. Note that the prepared Co/Ni multilayer samples exhibits perpendicular magnetic anisotropy (K$_{\mathrm{u}}$ = 662 kJ\,m$^{-3}$, saturation magnetization M$_{\mathrm{s}}$ = 940 kA\,m$^{-1}$) and is part of a sample series presented in Ref.~\cite{Hassan2024} where its magnetic properties are shown. 

\subsection{Lorentz transmission electron microscopy with \textit{in-situ} optical excitation} The measurements were performed at the Göttingen Ultrafast Transmission
Electron Microscope, which enables LTEM with \textit{in-situ} optical excitation~\cite{Feist2017, Eggebrecht2017, Moeller2020,Titze2024}. LTEM images were recorded using an acceleration voltage of 200\,keV at typical defocus values between 0.5 and 6\,mm utilizing a CEOS CEFID equipped with a TVIPS XF416 at a 10\,eV window width to provide energy filtering. The Co/Ni multilayer is excited by ultrashort laser-pulses (pulse duration of approximately 300\,fs) at a center wavelength of 750\,nm, close to normal angle of incidence, and a $1/e^2$ beam diameter of $60$\,{\textmu}m~\cite{Titze2024}. Snapshots of the magnetic spin textures were acquired before and after single-pulse excitation, varying both the oop magnetic field and the excitation fluence. The initial state spin textures before laser illumination were set by fully saturating the sample using an oop magnetic field of at least 141\,mT before approaching the respective magnetic field value. The field of view of the $4096 \times 4096$ pixel images is evaluated to be $21.3 \times 21.3$\,$\mu$m$^2$, i.e., one pixel has a size of $5.2 \times 5.2$\,nm$^2$.

\subsection{Data processing and analysis}
Starting from the single domain state, laser excitation is capable of nucleating a dense network of magnetic spin textures. A drastic quality improvement of LTEM images of these textures is obtained by subtracting the image of the initial single domain state, e.g., removing sample and camera artifacts. However, this procedure breaks down when starting from the stripe domain state and when tracking the evolution of the nucleated spin objects under increasing or decreasing oop magnetic fields. In particular, the initial state spin textures may disappear or transform upon laser excitation making their background subtraction impossible. Here, we make use of corresponding snapshots of the single domain state at different oop magnetic fields. Note that varying the oop magnetic field slightly modifies the imaging conditions in LTEM, primarily manifesting as rotations of the acquired images. To correct for these field-induced rotations, we track the evolution of three clearly distinguishable blisters throughout the field variation. 

Object detection for Fig.~\ref{fig:Nucleation} and Fig.~\ref{fig:Overview} was performed using a trained neural network based on \textit{yolov8}~\cite{yolov8}. Selected images at negative magnetic fields were tiled into 64 equal-sized parts ($512 \times 512$\,pixels), with each spin object individually labeled. A total of 1390 tiles were used as training data, containing approximately 8000 labeled objects, while 320 additional tiles were reserved for validation during the training process.

\begin{figure}[h!]
     \centering
     \includegraphics[width=1\columnwidth]{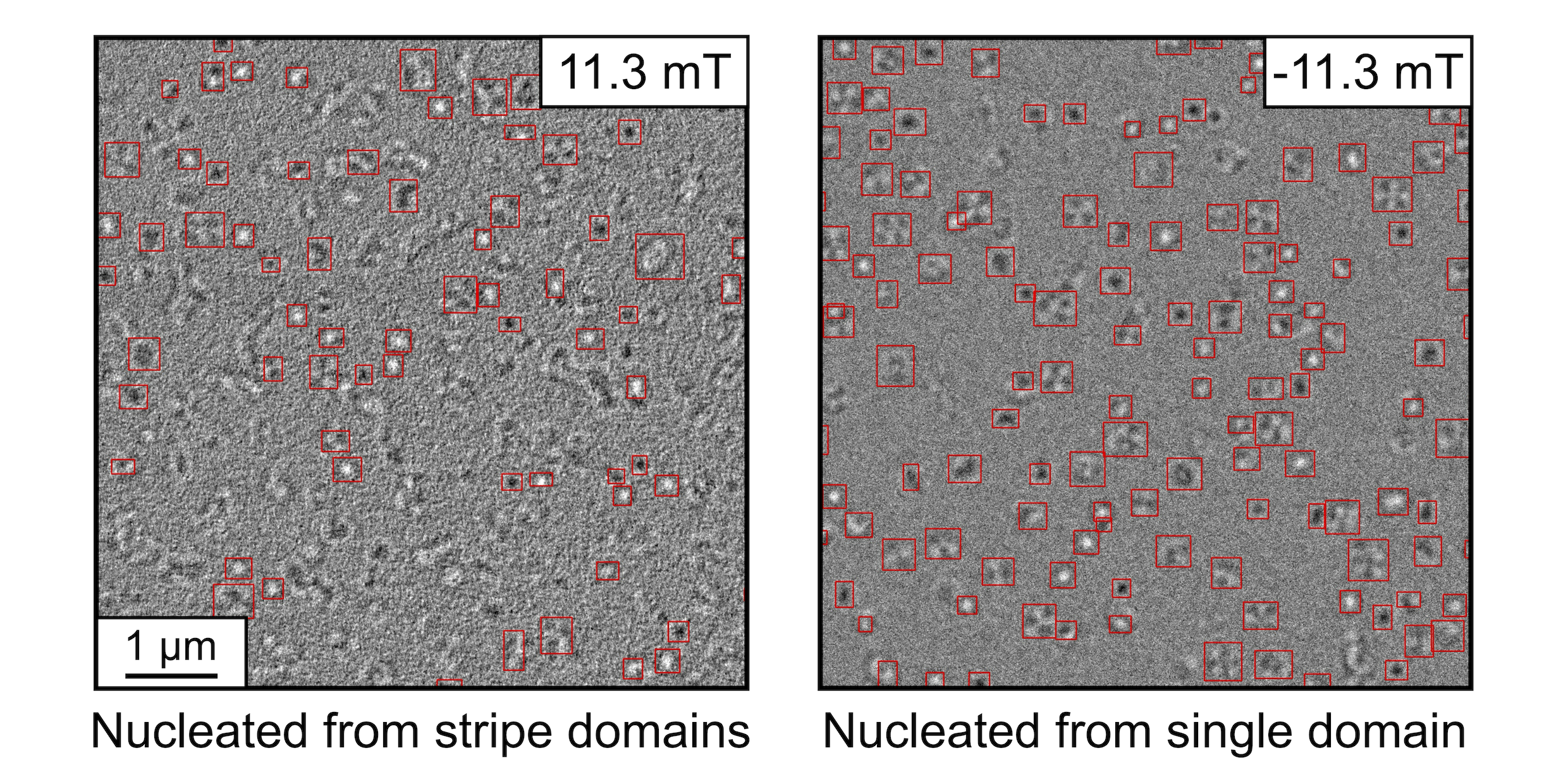}
     \caption{Red boxes mark SK/ASKs detected by the neural network laser-nucleated from a stripe domain (left) and single domain state (right).}
     \label{fig:Detection}
\end{figure}
Figure~\ref{fig:Detection} illustrates the differences in detection efficiency of skyrmions nucleated from a stripe domain state and a single domain state obtained after single excitation with fluence $F=4.8$\,mJ/cm$^2$ at an magnetic field of $\mu_0 H=\pm11.3$\,mT. Spin objects detected by the neural network are marked by red boxes in the two LTEM images, showcasing the decreased accuracy of the network when starting from a stripe domain initial state. 

After neural network detection, spurious detections are removed in a postprocessing step based on object geometry (size or aspect ratio, containment by larger spin structures) and intensity (predominantly black or white objects). This mainly affects the stripe domain states, where the image rotation for background subtraction leads to artifacts, particularly at images edges. Furthermore, an additional postprocessing script for the stripe domain initial state disregards spin objects of fixed $Q$, which are present at roughly the same spatial position both in the initial state image and the image after laser-excitation.

For visualization purposes only, the LTEM difference images shown in this work were Gaussian-filtered and contrast-adjusted. Calculated LTEM contrast (see below) was matched manually to the experimental data in Fig.~\ref{fig:Nucleation}.

\subsection{Calculated LTEM contrast from micromagnetic simulations}
Micromagnetic simulations were performed with the GPU-accelerated finite-difference framework \textit{magnum.np}~\cite{Bruckner2023}. Analogous to our previous work~\cite{Hassan2024}, we simulate a single layer ferromagnetic film, which was discretized into $1024\times1024\times1$ cells of size $2\times2\times\SI{4.5}{\nano\meter\cubed}$. Pseudo periodic boundary conditions were applied in both in-plane directions. We use a saturation magnetization $M_\mathrm{s}=\SI{940}{\kilo\ampere\per\meter}$, an exchange stiffness $A=\SI{10}{\pico\joule\per\meter}$, and a perpendicular uniaxial anisotropy
$K_\mathrm{u}=\SI{575}{\kilo\joule\per\meter\cubed}$ (easy axis along $z$). The effective field contains the exchange field, the uniaxial anisotropy field, the demagnetizing field, the external Zeeman field, and a stochastic thermal field at $T=\SI{300}{\kelvin}$ implemented as a Gaussian white-noise process. The magnetic states are obtained relaxing a random state by numerically solving the Landau-Lifshitz-Gilbert equation with a Gilbert damping constant $\alpha=1$. The system was relaxed independently for \SI{5}{\nano\second} at each value of a static out-of-plane field $\mu_0H_z$ ranging from $+100$ to $\SI{-100}{\milli\tesla}$ in steps of \SI{2}{\milli\tesla}. 

Fresnel-mode Lorentz transmission electron microscopy (LTEM) images were
computed from the relaxed magnetization states in two steps. First, the
magnetic phase shift that the sample imprints on the electron wave, is calculated. Then, we propagate this wave to the defocused image plane, where the phase becomes visible as an intensity contrast. 

For this purpose, an electron is passing through the film acquires an Aharonov-Bohm phase
$\varphi_m=(e/\hbar)\int A_z\,\mathrm{d}z$ set by the magnetic vector
potential $\mathbf{A}$ of the sample~\cite{Aharonov1959}. We evaluate
this phase in Fourier space, where the vector potential generated by the
magnetization has a simple closed form~\cite{Beleggia2003}:
\begin{equation}
\widetilde{\varphi}_m(\mathbf{q}_\perp)=
\frac{i\,e\,\mu_0}{\hbar}\,
\frac{\left[\widetilde{\mathbf{M}}_\perp(\mathbf{q}_\perp)\times\mathbf{q}_\perp\right]_z}{q_\perp^{2}},
\end{equation}
with $\mathbf{M}_\perp=\int\mathbf{M}\,\mathrm{d}z$ the
thickness-integrated magnetization. Because the film is
flat and of uniform thickness, the electrostatic phase is the same
everywhere and adds no contrast.

The magnetic textures are made visible by defocusing the microscope, which was
simulated by propagating the wave $\psi_0 = e^{i\varphi_m}$ over the defocus distance $\Delta f$ with
the standard contrast-transfer function~\cite{Degraef2003},
\begin{equation}
I(\mathbf{r}_\perp)=
\left|\mathcal{F}^{-1}\!\left\{
e^{-i\pi\lambda\,q_\perp^{2}\left(\Delta f+\frac{1}{2}C_\mathrm{s}\lambda^{2}q_\perp^{2}\right)}\,
\mathcal{F}\!\left[\psi_0\right]\right\}\right|^{2},
\end{equation}
where $\lambda=\SI{2.51}{\pico\meter}$ is the electron wavelength at the
simulated acceleration voltage of \SI{200}{\kilo\volt}. We assumed an
aberration-free ($C_\mathrm{s}=0$), fully coherent microscope and an
under-focus of $\Delta f=\SI{-0.7}{\milli\meter}$.

\bibliography{references}

\end{document}